%% file: main-with-bbl.tex
\documentclass[sigconf,natbib=true]{acmart}

\copyrightyear{2025}
\acmYear{2025}
\setcopyright{acmlicensed}\acmConference[SIGIR '25]{Proceedings of the 48th International ACM SIGIR Conference on Research and Development in Information Retrieval}{July 13--18, 2025}{Padua, Italy}
\acmBooktitle{Proceedings of the 48th International ACM SIGIR Conference on Research and Development in Information Retrieval (SIGIR '25), July 13--18, 2025, Padua, Italy}
\acmDOI{10.1145/3726302.3730343}
\acmISBN{979-8-4007-1592-1/2025/07}

\usepackage{natbib}
\usepackage{xspace}
\usepackage[simplified]{pgf-umlcd}
\usepackage{float}
\usepackage{listings}
\usepackage{listingsutf8}
\usepackage{multirow}
\usepackage{subcaption}
\usepackage{graphicx}
\usepackage{fixltx2e}
\usepackage{soul}
\usepackage{cancel}
\usepackage[absolute,overlay]{textpos} 
\usepackage{todonotes}

\definecolor{mediumgray}{gray}{0.60}
\definecolor{webiscodebasic}{rgb}{0.2,0.2,0.2}
\definecolor{webiscodekeyword}{rgb}{0.0,0.5,0.0}
\definecolor{webiscodekeywordself}{rgb}{0.7,0.4,0.6}
\definecolor{webiscodeidentifier}{rgb}{0.0,0.0,0.0}
\definecolor{webiscodecomment}{rgb}{0.25,0.5,0.5}
\definecolor{webiscodestring}{rgb}{0.75,0.12,0.12}
\definecolor{webiscodedecorator}{rgb}{0.6,0.3,0.0}
\lstdefinestyle{webisstyle}{
  basicstyle=\fontsize{7.45pt}{8.45pt}\selectfont\ttfamily\color{webiscodebasic},
  keywordstyle=\color{webiscodekeyword},
  identifierstyle=\color{webiscodeidentifier},
  commentstyle=\color{webiscodecomment},
  stringstyle=\color{webiscodestring},
  showstringspaces=false,
  frame=lines,
  framesep=0.7em,
  rulesep=0.5em,
  framerule=0.1em,
  keepspaces=true,
  tabsize=2,
  showtabs=false,
  numbers=none,
  literate=
    {->}{{\textrightarrow}}{2}
    {>=}{{\(\geq\)}}{2}
    {<=}{{\(\leq\)}}{2}
    {!=}{{\(\neq\)}}{2},
}
\newfloat{listing}{htbp}{lop}
\floatname{listing}{Listing}

\begin{document}
\pagenumbering{gobble}
\title{\texttt{ir\_explain}: a Python Library of Explainable IR Methods}

\newcommand{\irexplain}{\texttt{\small ir\_explain}\xspace}
\newcommand{\iraxioms}{\texttt{\small ir\_axioms}\xspace}
\newcommand{\irdatasets}{\texttt{\small ir\_datasets}\xspace}
\newcommand{\irmeasures}{\texttt{\small ir\_measures}\xspace}
\newcommand{\module}[1]{{\texttt{\small #1}}\xspace}

\author{Sourav Saha}
\affiliation{%
	\institution{Indian Statistical Institute, Kolkata}
	\city{}
	\country{India}
}
\email{sourav.saha\_r@isical.ac.in}

\author{Harsh Agarwal}
\affiliation{%
	\institution{Indian Statistical Institute, Kolkata}
	\city{}
	\country{India}
}
\email{1996harsh.agarwal@gmail.com}

\author{Venktesh V}
\affiliation{%
	\institution{TU Delft}
	\city{}
	\country{The Netherlands}
}
\email{V.Viswanathan-1@tudelft.nl}

\author{Avishek Anand}
\affiliation{%
	\institution{TU Delft}
	\city{}
	\country{The Netherlands}
}
\email{Avishek.Anand@tudelft.nl}

\author{Swastik Mohanty}
\affiliation{%
	\institution{Indian Statistical Institute, Kolkata}
	\city{}
	\country{India}
}
\email{swastik.mohanty63@gmail.com}

\author{Debapriyo Majumdar}
\affiliation{%
	\institution{Indian Statistical Institute, Kolkata}
	\city{}
	\country{India}
}
\email{debapriyo@isical.ac.in}

\author{Mandar Mitra}
\affiliation{%
	\institution{Indian Statistical Institute, Kolkata}
	\city{}
	\country{India}
}
\email{mandar@isical.ac.in}

\begin{abstract}
\input{abstract}

\end{abstract}

%
%
\settopmatter{printacmref=true, printccs=true, printfolios=true}


\maketitle

\input{intro}
\input{explainable-ir}
\input{irexplain-library}
\input{experiments}
\section*{Acknowledgement}
We thank the anonymous reviewers for comments on earlier versions of this draft, which resulted in significant improvements in the current version. This work is partially supported by German Research Foundation (DFG), under the Project IREM with grant No. AN 996/1-1. 
Sourav Saha is supported by TCS Research Scholar program (Cycle 17). He also acknowledges Sobigdata++ team for a support to visit TU Delft. 


\end{document}

%% file: abstract.tex

While recent advancements in Neural Ranking Models have resulted in
significant improvements over traditional statistical retrieval models, it
is generally acknowledged that the use of large neural architectures and
the application of complex language models in Information Retrieval (IR)
have reduced the transparency of retrieval methods. Consequently,
Explainability and Interpretability have emerged as important research
topics in IR. Several axiomatic and post-hoc explanation methods, as well
as approaches that attempt to be interpretable-by-design, have been
proposed. We present \irexplain, an open-source Python library that
implements a variety of well-known techniques for Explainable IR (ExIR)
within a common, extensible framework. It supports the three standard
categories of post-hoc explanations, namely pointwise, pairwise, and
listwise explanations. The library is designed to make it easy to reproduce
state-of-the-art ExIR baselines on standard test collections, as well as to
explore new approaches to explaining IR models and methods. To facilitate
adoption, \irexplain is well-integrated with widely-used toolkits such as
Pyserini, PyTerrier (work in progress) and \irdatasets. Downstream
applications of \irexplain include explaining the Retrieval-Augmented
Generation (RAG) pipeline. The development version of the library is
available on GitHub. We release the library as a pip package
(\url{https://pypi.org/project/ir-explain/}); source code is available from
\url{https://github.com/souravsaha/ir_explain}.

%% file: intro.tex
\section{Introduction and motivation}
\label{sec:intro}
Large Language Models (LLMs)~\cite{devlin2018bert, raffel2023exploring,
  gpt3, ouyang2022training, chowdhery2022palm, touvron2023llama} are the
most recent in a succession of increasingly complex deep learning models
that have been applied to Information Retrieval (IR). These models are
frequently almost magically effective at fulfilling users' information
needs in a wide variety of tasks. At the same time, the models are also
known to have important flaws, e.g., a tendency to `hallucinate' in certain
circumstances. If the output of such a system included (in addition to the
final results) an \emph{explanation} of how the system arrived at these
results, an end-user would be better able to judge the trustworthiness of
the answer. This would likely lead to increased adoption of these
remarkable technologies in high-risk situations. Accordingly, the emphasis
on `understanding' these complex models has increased, and a good deal
of research has been done in recent times on 
the explainability of Machine Learning (ML) models in general, and the
so-called `mechanistic interpretability' of LLMs in particular. Our focus
in this work is on methods that are specific to Explainable IR (ExIR). While
these methods are often inspired by ML explainability techniques, they have
a distinct flavour and purpose because they explicitly deal with ranking
rather than classification. A detailed overview of research in the broad
area of ExIR can be found in~\cite{anand2022explainable,
  saha2022explainability, exir_tutorial_sigir}. The diverse research
efforts summarised in these overviews constitute a good start, but more
concerted attention is required from the community in order to address the
many challenges that remain, before IR systems can explain their output in
ways that are simultaneously precise, clear and intuitive.

\subsection*{Our contribution}
In order to encourage new research in this area, we have put together
\irexplain, a Python library of ExIR methods. To the best of our knowledge,
this is the first library that implements all the major approaches that
have been proposed so far to explain rankings produced by complex IR
models. The target audience of \irexplain includes researchers working in
ExIR who are looking for a convenient way to compare newly proposed
approaches with existing ones, as well as practitioners who
simply want to applying existing methods to `debug' or explain neural
ranking models (NRMs) that are used in practical
applications. 
We hope this package will also reduce the barrier to entry for new
researchers, and make it easy for them to explore all major existing
ExIR techniques, and experiment with new ones.

\irexplain is intended to be well-integrated with widely-used toolkits such
as Pyserini and PyTerrier, as well as \irdatasets for test collection
management. The library is thus expected to eliminate the burden of
(i)~locating available ExIR implementations scattered across different
locations, and (ii)~setting them up to work with one's choice of an engine
for running IR experiments. 

\irexplain provides a framework for implementing and analysing new
approaches to ExIR, and includes an evaluation component for measuring the
fidelity of explanations via similarity measures commonly used for
comparing IR rankings.

To showcase its capabilities, we use \irexplain in this study to (i)
analyse the robustness of a pointwise explanation method, (ii) examine
whether the explanations (lists of terms) produced by listwise methods are
sufficiently intuitive, (iii) provide a usecase of pairwise explanation, (iv) study the replicability of
previously reported experimental results obtained using listwise explainers, and (v) demonstrate \irexplain in the retrieval-augmented generation (RAG) pipeline.
Source code for \irexplain is available from
\url{https://github.com/souravsaha/ir_explain}. \irexplain is also available for
installation from the Python Package Index (\url{https://pypi.org/ir-explain/}) using 
\module{pip install ir-explain}.

\section{Related work}
The IR community has a long tradition of providing open-source libraries
and resources that make it convenient for researchers to experiment with,
evaluate and understand different retrieval models.
Anserini~\cite{anserini_sigir_2017},
Pyserini~\cite{Lin_etal_SIGIR2021_Pyserini},
PyTerrier~\cite{pyterrier2020ictir} and
PyGaggle~\cite{pygaggle-nogueira-etal-2020-document} are examples of modern
IR toolkits that support both sparse and dense retrieval models, and
provide easy access to the standard retrieve-and-rerank pipeline on TREC
collections.

Other well-known IR resources that are related to \irexplain include
\irdatasets~\cite{ir_dataset_sigir_21} (for acquiring and managing test
collections for ad hoc IR) and \iraxioms~\cite{bondarenko:2022d}. The
\iraxioms package implements various retrieval axioms~\cite{fang-sigir-04},
as well as a method for aggregating the document ordering preferences
specified by different axioms into a single
ranking~\cite{hagen:2016:cikm:aximaticReranking}. To facilitate working
with standard datasets and IR tools, \iraxioms is tightly integrated with
\irdatasets and PyTerrier. The design of \irexplain is strongly influenced
by \iraxioms. Indeed, \irexplain makes heavy use of \iraxioms in order to
provide axiom-based explanations. Like\linebreak{}\iraxioms, \irexplain is also
tightly integrated with \irdatasets. However, while \iraxioms is integrated
with PyTerrier, the current version of \irexplain makes use of
Pyserini to access indexed collections. Integrating \irexplain with indices
created by PyTerrier is work in progress.

While implementations of well-known ExIR approaches are not currently
available within a single package, high quality interpretability libraries
for ML models do exist.
The AIX360 toolkit from IBM~\cite{aix360-jmlr,aix360} implements ``diverse
state-of-the-art explainability methods [including both global and local
post-hoc explanations (e.g., LIME and SHAP), as well as `direct'
explanations], two evaluation metrics,
and an extensible software architecture''. Its primary focus is on binary
classification, though regression is also addressed via Generalized
Linear Rule Models~\cite{wei2019generalized}.

Captum (\url{https://captum.ai} also implements many of the well-known
interpretability techniques for ML models used in Computer Vision and Text
Classification, e.g., feature attribution based on simple feature ablation,
Integrated Gradients~\cite{sundararajan:2017:integratedgradients} or
\linebreak{}DeepLIFT~\cite{shrikumar:2017:icml:deeplift}, and LIME~\cite{lime}. Captum
is tightly integrated with PyTorch. While this provides easy access via
\texttt{torchtext.datasets} to textual datasets like AG's News Corpus
(\cite{character-cnn-neurips-2015}, also see
\url{http://www.di.unipi.it/˜gulli/AG_corpus_of_news_articles.html}) and
the IMDb Large Movie Review Dataset~\cite{imdb-large-movie-review-dataset},
these collections are intended for tasks like text / sentiment
classification, machine translation, question answering, and sequence
tagging, rather than ranking and retrieval.
Given the classification-oriented nature of AIX360 or Captum, and the level
of integration between \irexplain and existing IR resources, we believe
\irexplain is a much more natural fit for research and development in ExIR.


%% file: explainable-ir.tex
\section{\texorpdfstring{\texttt{\LARGE\lowercase{ir\_explain}}}{IR
    Explain}: background and design}
A two-stage pipeline is one of the standard paradigms used in modern Neural
IR. 
First, given a query $Q$, a sparse retrieval
model (e.g., BM25 or language models with Jelinek-Mercer (LMJM) or Dirichlet (LMDir) smoothing) 
is used to initially
retrieve top $k$ documents. 
The number $k$ is chosen to be small compared to the collection size. 
In the second stage, a neural matching model (e.g., a cross-encoder or a dual-encoder) is employed to re-rank the set of $k$
initially retrieved documents. 
Let us denote the first-stage ranker by $M_1$ and the second-stage ranker by $M_2$. 
The central problem of ExIR is to explain different aspects of these two stages of ranking.

\begin{figure*}[t]
  \centering
  \includegraphics[scale=0.7]{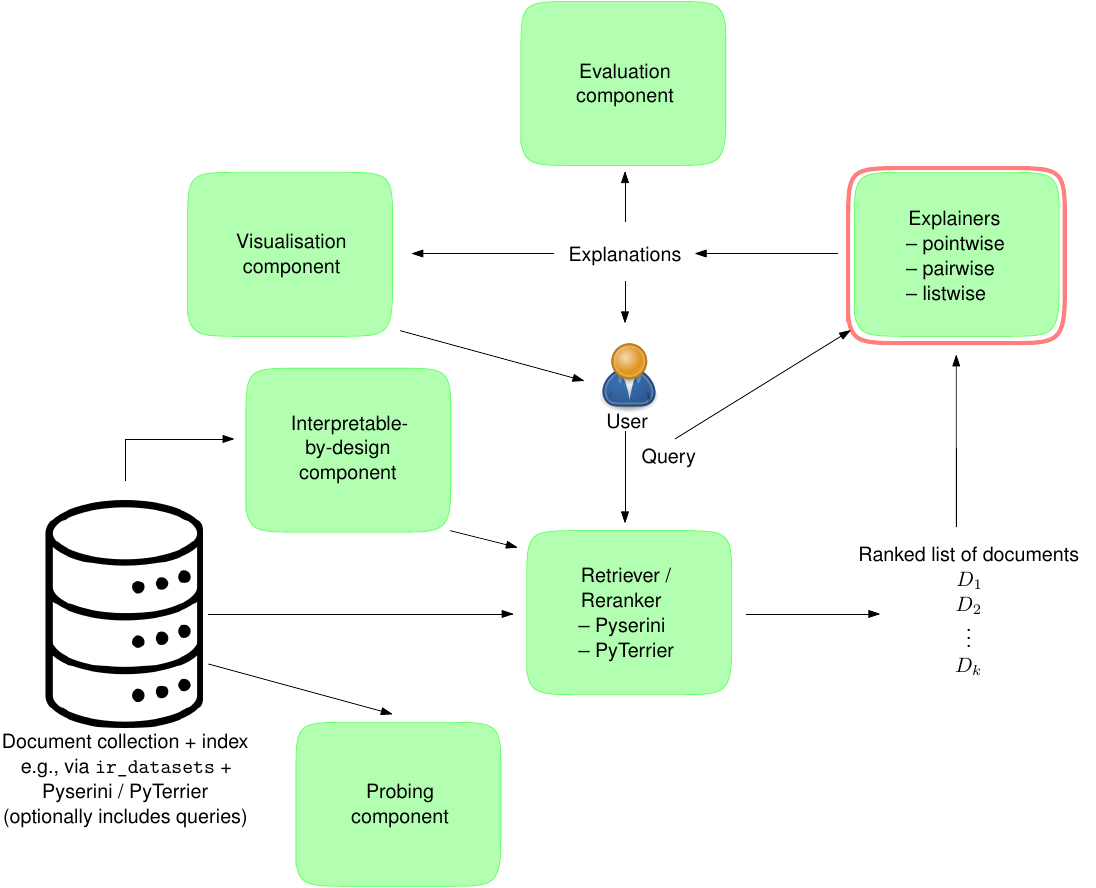}
  \caption{The overall structure of \irexplain showing how the various
    components of the library are intended to be used in practice.}
  \label{fig:schematic}
\end{figure*}

\input{inheritance}

While traditional, sparse retrieval models are generally regarded as
\emph{interpretable by design}, \emph{post-hoc explanations} are most
common in the case of neural ranking models (NRMs). These explanations may
be classified into two groups based on whether they address (a)~the
document representation that is used to compute scores, or (b)~the ranking
decision induced by the model. Document representations generated by neural
models are usually explained using \emph{probing} methods, which study the
lexical or semantic features that are encoded within these representations.
Explanations of ranking decisions are in turn categorised as (i) pointwise,
(ii) pairwise, and (iii) listwise explanations. 

Figure~\ref{fig:schematic} presents the overall structure of \irexplain,
showing components that offer the various kinds of explanations listed
above, along with auxiliary, supporting components. We briefly discuss the
interpretable-by-design and probing components of \irexplain in
Sections~\ref{sec:ibd} and \ref{sec:probe}. In the rest of this section,
we describe the core component of \irexplain (highlighted in
Figure~\ref{fig:schematic}), consisting of three major sub-modules, which
implement pointwise, pairwise and listwise explanation methods.
These sub-modules 
are designed so that they may be used to explain ranked lists generated on
the fly (by specifying only a dataset, and any one of the dense retrieval,
learned sparse retrieval, or hybrid retrieval models supported by Pyserini
or PyTerrier). They may also be used to explain \emph{any} ranked list
(\module{runfile}) generated by 
leveraging an IR engine of one's choice, if such a list is supplied as an
input to \irexplain via the \module{load\_from\_res}
function. 
%
%

\subsection{Pointwise Explanations}
\label{sec:point-wise}
In general, a pointwise explanation attempts to address two questions : (i) why is a particular document $D$ retrieved within the top ranks by $M_1$? 
(ii) why does reranking by $M_2$ 
improve (or adversely affect) the rank of $D$? 
The explanations are typically in the form of a list of terms, along with
weights that indicate how these terms influence the ranking
of $D$, positively or negatively. Representative pointwise approaches
include EXS~\cite{singh:2019:wsdm:exs} and
LIRME~\cite{verma:2019:sigir:lirme}, which uses ideas from
LIME~\cite{lime} within a document-ranking setup.

Figure~\ref{fig:inheritance-fig} shows the inheritance relation
between the various classes of explainers implemented in \irexplain.
\linebreak\module{BasePointwiseExplainer} inherits from the \module{BaseExplainer}
class; in turn, \module{LirmePointwiseExplainer} and
\module{EXSPointwiseExplainer} classes inherit from
\module{BasePointwiseExplainer} class. One can easily add a new point-wise
explanation approach by inheriting from this base class, and overloading
the \module{explain} function. Listing~\ref{listing-pointwise-explanation}
demonstrates how LIRME may be used to explain a particular {query-document}
pair. The LIRME explainer is instantiated by passing it the index for a
collection $C$; additional parameters such as \module{kernel},
\module{sampling\_method}, and the number of desired explanation terms can
be set through a dictionary. The \module{explain()} function of the
\module{LirmePointwiseExplainer} class takes a query $Q$, a document $D$
and additional parameters as input, and returns, as an \emph{explanation
  vector}, a set of terms
and their contributions (positive or negative) to the rank of $D$. 

The \module{TermVisualization} class can be
instantiated to store an explanation vector, and its
\module{visualize()} function helps visualize the vector, as shown in Figure~\ref{fig:subfigures}. The \module{eval} module of \irexplain helps to evaluate different types of explanations. \module{PointWiseCorrectness} and \module{PointWiseConsistency} classes and their \module{evaluate()} methods
implement the ground-truth generation strategies proposed in LIRME.
All the different sampling strategies proposed in both LIRME and EXS have been included.
\begin{listing}[tb]
\begin{lstlisting}[
        firstline=2,
    ]
# set parameter for LIRME
params = {
  "sampling_method" : "random",   # sampling method
  "top-terms" : 10,               # explanation terms
  ...
}

# instantiate object of LIRME explainer
lirme = LirmePointwiseExplainer(model, corpus_path = 
    "/path/to/index", indexer_type= "pyserini")

# generate explanation vector
explan_vector, ranked_list = lirme.explain(query,
    docid, params)
    
# visualize explanations
visualizer = TermVisualization()
visualizer.visualize(explan_vector, show_top=5) 
    
# Evaluation component
# compute pointwise explanation correctness 
correctness = PointWiseCorrectness(lirme)
correctness.evaluate(query, docid, explan_vector)

# compute pointwise explanation consistency
consistency = PointWiseConsistency(lirme)
consistency.evaluate(query, docid, ranked_list)
\end{lstlisting}
    \caption{Example of a pointwise explainer LIRME with a query, document, and a set of parameters passed to it.}
    \label{listing-pointwise-explanation}
\end{listing}
\subsection{Pairwise Explanations}
\label{sec:pair-wise}
A pairwise explanation attempts to explain why a document $D_i$ is
preferred over another document $D_j$ for a particular query $Q$. Such
explanations are  often provided in terms of retrieval axioms. 
Axioms are formalisations of intuitive retrieval heuristics that specify constraints that a good ranking method should fulfil. 
Several sets of axioms have been formulated in the literature~\cite{fang:2005:sigir:axiomaticIR,amigo:2017:sigir:axiomaticIR, fang-sigir-04}, and a few relaxations have been proposed to make the axioms usable in practice \cite{hagen:2016:cikm:aximaticReranking}. 
\irexplain leverages the \iraxioms library for pairwise explanations.
Following \iraxioms, each retrieval axiom corresponds to a separate Python
class in \irexplain.\footnote{The \module{Argumentativeness} axioms have
  been omitted for now, since they do not apply to the ad-hoc retrieval
  settings used in our experiments.} 
Recall from~\cite{bondarenko:2022d} that, given
$\langle Q, D_i, D_j \rangle$, an object corresponding to axiom $A$ returns a
preference score $\mathit{pref}_{A}(Q,D_i,D_j) \in \{-1,0,+1\}$ that
specifies whether $D_i$ is preferred over $D_j$ by $A$: a score of $+1$
(resp.\ $-1$) signifies $D_i$ is (resp.\ not) preferred over $D_j$, and a $0$
means no preference.
The implementation in \iraxioms is, however, tightly integrated with the Pyterrier retrieval pipeline, and does not seem to have a provision for 
explaining an arbitrary document pair $\langle D_1, D_2 \rangle$;
this feature has been included in \irexplain. As in \iraxioms, users can use binary and unary operators to aggregate different axioms, and may also easily define new axioms.
In \irexplain, we have segregated all the axioms into a separate pairwise explanation module.

Listing~\ref{listing-pairwise-explanation} demonstrates how this module
works in \irexplain. The \module{PairwiseAxiomaticExplainer} class is instantiated with $\langle Q, D_i, D_j \rangle$
and the path to a collection index. This class is inherited from the \module{BaseExplainer} class.
The \module{explain()} function takes a list of \module{axioms} for which the preference needs to be computed. 
For more involved axioms (such as \module{PROX1}, \module{PROX2}, 
\module{PROX3}~\cite{bondarenko:2022d}), the
preference score alone does not adequately explain why one document is
preferred over another. The function \module{explain\_details()} (new in \irexplain) may be
used to obtain a more detailed understanding in such cases. For example, according to the \module{PROX1} axiom, a document $D_1$ is preferred over $D_2$ if $D_1$ has a shorter average distance between the query terms. Table~\ref{table-explain-detail-func} shows the output of \module{explain\_details()} for $\langle Q, D_i, D_j \rangle$, 
which compares the raw term frequencies (\module{tf}) of query terms, as
well as the average distance between the query tokens in $D_i$ and $D_j$.
\begin{listing}[tb]
\begin{lstlisting}[
        firstline=2,
    ]
# instantiate pairwise object
pairwise = PairwiseAxiomaticExplainer(query, doc1, 
    doc2, index_path)

# instances of various axiom classes
axiom_classes = [TFC1(), STMC1(), ...]

# call the explain method
pairwise.explain(axiom_classes)

# call the explain_details method to understand further
pariwise.explain_details(query, doc1, doc2, axiomName)
\end{lstlisting}
    \caption{Example use-case of a pairwise explanation module to explain a query-document pair. A list of axiom objects can be passed to the explain function.
    }
    \label{listing-pairwise-explanation}
\end{listing}

\begin{table}[tb]
    \centering
    \caption{Output of \module{explain\_details} method 
    for the query \emph{`exons definition biology' (qid: 183378)} and a pair of documents (\module{D1077802 and D1806793}). These two documents are sampled from the MS MARCO document collection.
    }
    \label{table-explain-detail-func}
    \fontsize{8pt}{10pt}\selectfont
    \renewcommand{\tabcolsep}{6pt}
    \renewcommand{\arraystretch}{1}
    \begin{tabular}{@{}cccc@{}}
        \toprule
        \multicolumn{3}{c@{}}{\textit{Query: exons definition biology (qid: 183378) }} \\
        \midrule
        \module{docid} & \module{D1077802} & \module{D1806793} \\
        \midrule
        \module{tf(exon)} & 23	& 21 \\
        \module{tf(definit)} & 7 &	56 \\
        \module{tf(biolog)} & 1	& 25 \\
        \module{avg\_dist(exon, definit)}  & 174.43	 &  2728.07 \\
        \module{avg\_dist(definit, biolog)}  & 354.71	&  3287.24\\
        \module{avg\_dist(exon, biolog)}  & 315.04	&  2864.24 \\
        \module{num pairs} & 3 & 3 \\
        \midrule
        \module{Total\_avg\_dist} & 281.39	&  2959.85 \\    
        \bottomrule
    \end{tabular}
\end{table}
\subsection{Listwise Explanations}
\label{sec:list-wise}
Listwise approaches provide a single explanation for an entire ranked list $L= \{D_1, D_2,..., D_k\}$.
These explanations are typically in the form of a pair $\langle Q_{\mathit{exp}},
\mathit{SM} \rangle$, where $Q_{\mathit{exp}}$ is an expanded version of the
query $Q$, and $\mathit{SM}$ is a `simpler' ranker, such as BM25 that is used to approximate a neural model.
The interpretation of this explanation is that, for $Q$, the complex ranker $M_2$ behaves as if it implicitly finds the same matches that $\mathit{SM}$ would explicitly find, given $Q_{\mathit{exp}}$.
The task of finding an explanation reduces to (i)~constructing
$Q_{\mathit{exp}}$, and (ii)~verifying that 
the ranked list generated using $\langle Q_{\mathit{exp}}, \mathit{SM} \rangle$ is approximately equivalent to $L$. Based on how these sub-tasks are accomplished, listwise
approaches can be divided into two broad categories. 
The first type considers $L$ 
as a combination of multiple ordered pairs of documents, and seeks to
explain as many of these pairs as possible. Thus, a list $\mathcal{L} = \{(D_j > D_k)
: D_j, D_k \in L\}$ of preference pairs is first constructed by sampling
document pairs from $L$. The next objective is to construct a
$Q_{\mathit{exp}}$, such that $\langle Q_{\mathit{exp}}, \mathit{SM} \rangle$ preserves the ordering of most pairs of $\mathcal{L}$.
The underlying hypothesis is that if $\mathit{SM}$ agrees with $M_2$ on most of the randomly selected preference pairs, it is consistent with the full ranked list. Multiplex~\cite{lijun-ecir-23} and IntentEXS~\cite{intent_exs_singh_fat_20} belong to this category.
The second class of approaches, exemplified by BFS~\cite{llordes-sigir-23}
and Greedy~\cite{llordes-sigir-23}, directly measure the similarity  
between $L$ and the list returned using $\langle Q_{\mathit{exp}}, \mathit{SM} \rangle$, without decomposing $L$ into pairwise preferences.
    
\irexplain implements four state-of-the-art listwise methods:
\begin{itemize}
    \item \module{MultiplexListwiseExplainer}, 
    \item \module{IntentListwiseExplainer}, 
    \item \module{BFSListwiseExplainer}, 
    \item \module{GreedyListwiseExplainer}.
\end{itemize} 
Each of the corresponding classes is inherited from the \linebreak\module{BaseListwiseExplainer} class (see Figure~\ref{fig:inheritance-fig}). 
Listing~\ref{listing-listwise-explanation} shows an example of how \module{MultiplexListwiseExplainer} can
be used to obtain an explanation. The \module{multiplex} object is an instance of the corresponding listwise class. Parameters such as the type of the simple explainer, document pair sampling strategy, 
optimization method, etc.\ are set via \module{dictionary}.
The \module{explain()} function internally invokes several stages of the
Multiplex explainer. We also support batch processing with the
\module{explain\_all()} function, which calls the \module{explain()} method
for all the queries in a specified dataset.
Additionally, \irexplain provides a few utility functions, e.g.,
\module{generate\_candidates()}, and\linebreak\module{show\_matrix()}, that provide white-box views into various stages of the \module{explain()} method. 
Multiplex uses three simple rankers as explainers, whereas IntentEXS uses only one to explain a ranked list. 
\irexplain provides an opportunity to use IntentEXS with any one of the three simple explainers used in Multiplex. 
We have also augmented the list of simple explainers to include other
widely used statistical retrieval models (such as LMDir and LMJM) in
addition to the single model (BM25) 
used for experiments by \citet{lijun-ecir-23}.
{Users can vary parameters
such as the strategy for sampling document pairs, 
and there is a provision to use pluggable components for some parameters. Thus, a custom helper routine may be used to sample document pairs based on some specific distribution.
\irexplain makes it convenient for the user to mix and match different explainers. 
One such use case is to adapt Multiplex 
in the framework of BFS. Further, in-built evaluation measures (briefly described in Section~\ref{sec:evaluation}) such as Rank-Biased Overlap (RBO~\cite{rbo}) {between the complex and simple ranker}
can be computed to compare the fidelity of the explanations provided by this variant of Multiplex, with baselines like BFS and Greedy. 

\begin{listing}[tb]
\begin{lstlisting}[
        firstline=2,
    ]
# set parameters for Multiplex
params = {
  "max_k" : 10,                    # max exp. terms
  "min_k" : 4,                     # max exp. terms
  "EXP_model" : "language_model",  # explainer type
  "optimize_method" : "geno",      # optim. method
  ...
  }
  
# instantiate multiplex and call explain function
multiplex = MultiplexListwiseExplainer(index_path = 
    "/path/to/index", indexer_type = "pyserini")

# invoke the explain method   
multiplex.explain(qid, query_str, params)
\end{lstlisting}
    \caption{Example of a listwise explainer Multiplex with a query, ranked list, and a set of parameters passed to the explainer.}
    \label{listing-listwise-explanation}
\end{listing}





\subsection{Interpretable by Design}
\label{sec:ibd}
Although these approaches are not as popular as post-hoc explanations, they
attempt to construct models that are, in principle, explainable by design.
Explanation methods of this type generally require retraining the ranking
model with interpretability in mind. SELECT-AND-RANK~\cite{select-and-rank}
is a recently proposed method from this category. It first extracts a small
subset of the sentences from each document, and ranks documents based
solely on these extracted snippets. Thus, the explanations provided in this
paradigm are essentially concise (and presumably more easily understood)
`summaries' of the documents. The proposed algorithms train the model in an
end-to-end fashion. \irexplain provides an interpretable-by-design
component to build a ranking model from scratch using this approach.
Additionally, a pretrained model will be made available for easy adoption of this approach.

\subsection{Probing}
\label{sec:probe}
In this module of \irexplain, the main emphasis is on understanding the
encoded representation of documents holistically. Specifically, probing
components allow us to analyze what IR properties (such as lexical and
semantic matching, etc.) are encoded in retrieval models. There are some
ongoing research efforts in this direction. \citet{causal-probing} proposed
a causal probing framework for a widely used dual encoder model to analyze
its capability of capturing six properties, namely: semantic matching,
lexical matching (BM25), coreference resolution, named entity recognition,
question classification, and term matching. \irexplain implements this
causal probing framework to facilitate extensive analysis of the different
layers of dual encoders and the {features} encoded within them.


\subsection{Evaluation}
\label{sec:evaluation}
Evaluation metrics related to ExIR, such as correctness, consistency, and
fidelity, are implemented in the \module{evaluation} module of \irexplain.
As illustrated in Listing~\ref{listing-pointwise-explanation}, the
\module{PointWiseCorrectness} class measures how well the pointwise
explanation terms correlate with the expanded terms sampled using LMJM. On
the other hand, the consistency measure (\module{PointWiseConsistency})
quantifies the relative differences in explanation terms when different
sampling methods are used. A more detailed discussion can be found
in~\cite{verma:2019:sigir:lirme, saha2022explainability}. 
For listwise explanations, metrics such as Jaccard similarity and RBO are used to measure the overlap between the ranked lists produced by a complex ranker and a simpler one. Intuitively, RBO quantifies the similarity between these two ranked lists, and attaches greater importance to commonalities at top ranks.
The
evaluation of IR explanations is constantly evolving, however; these modules will
need to be updated as newer and more robust evaluation approaches
emerge.


\subsection{Visualization}
\label{sec:visualization}
We provide visualization in the form of raw terms and their contributions (see Figure~\ref{fig:subfigures}) for now. Many general-purpose visuzalization tools are available, but our current focus is primarily on explaining retrieval models. However, we plan to integrate \irexplain with various visualization tools and support different output formats such as \module{html}, \module{json}, etc. 





\subsection{Utilities}
\label{sec:utils}

We discuss several utilities provided by \irexplain. One such popular component is document perturbation. There are different strategies to perturb a document and generate the perturbed instance $D'$. We provide \module{RandomSampler}, \module{MaskingSampler}, and \module{TfIdfSampler}. \module{RandomSampler} is the simplest; it randomly removes words from a document. In contrast, \module{TfIdfSampler} samples words from a document based on their tf-idf weights. A more sophisticated sampling based techniques can be added easily by using these utils module.

%% file: inheritance.tex
\begin{figure*}[ht]
    \begin{tikzpicture} 
    
        \begin {class}[text width = 4 cm]{BaseExplainer}{7, 2}
            \operation{explain()}
            \operation{preprocess()}
        \end {class}
        
        \begin {class}[text width = 4 cm]{BasePointwiseExplainer}{1, 0}
            \inherit{BaseExplainer}
            \operation{explain($Q$, $D$)}
        \end {class}
    
        \begin {class}[text width = 3.5 cm]{LirmePointwiseExplainer}{0, -1.5}
            \inherit{BasePointwiseExplainer}
            \operation{explain($Q$, $D$)}
        \end {class}
    
        \begin {class}[text width = 3.5 cm]{EXSPointwiseExplainer}{4, -1.5}
            \inherit{BasePointwiseExplainer}
            \operation{explain($Q$, $D$)}
        \end {class}
        
        \begin {class}[text width = 4 cm]{PairwiseAxiomaticExplainer}{7 , 0}
            \inherit{BaseExplainer}
            \operation{explain($Q$, $D_j$, $D_k$)}
        \end {class}
    
        \begin {class}[text width = 4 cm]{BaseListwiseExplainer}{12, 0}
            \inherit{BaseExplainer}
            \operation{explain($D_1, ..., D_k$)}
        \end {class}

        \begin {class}[text width = 3.5 cm]{BFSListwiseExplainer}{3.5, -3}
            \inherit{BaseListwiseExplainer}
            \operation{explain($Q$, $D_1, ..., D_k$)}
        \end {class}

        \begin {class}[text width = 3.5 cm]{GreedyListwiseExplainer}{8, -3}
            \inherit{BaseListwiseExplainer}
            \operation{explain($D_1, ..., D_k$)}
        \end {class}

        \begin {class}[text width = 3.5 cm]{IntentListwiseExplainer}{9.5, -4.5}
            \inherit{BaseListwiseExplainer}
            \operation{explain($D_1, ..., D_k$)}
        \end {class}

        \begin {class}[text width = 4 cm]{MultiplexListwiseExplainer}{13.6, -4.5}
            \inherit{BaseListwiseExplainer}
            \operation{explain($D_1, ..., D_k$)}
        \end {class}

    \end{tikzpicture}
    \caption{The inheritance class diagram of the different explanation modules used in \irexplain for explaining ranking decisions.}
    \label{fig:inheritance-fig}

\end{figure*}

%% file: experiments.tex
\begin{figure*}
    \centering
    \includegraphics[width=0.9\linewidth]{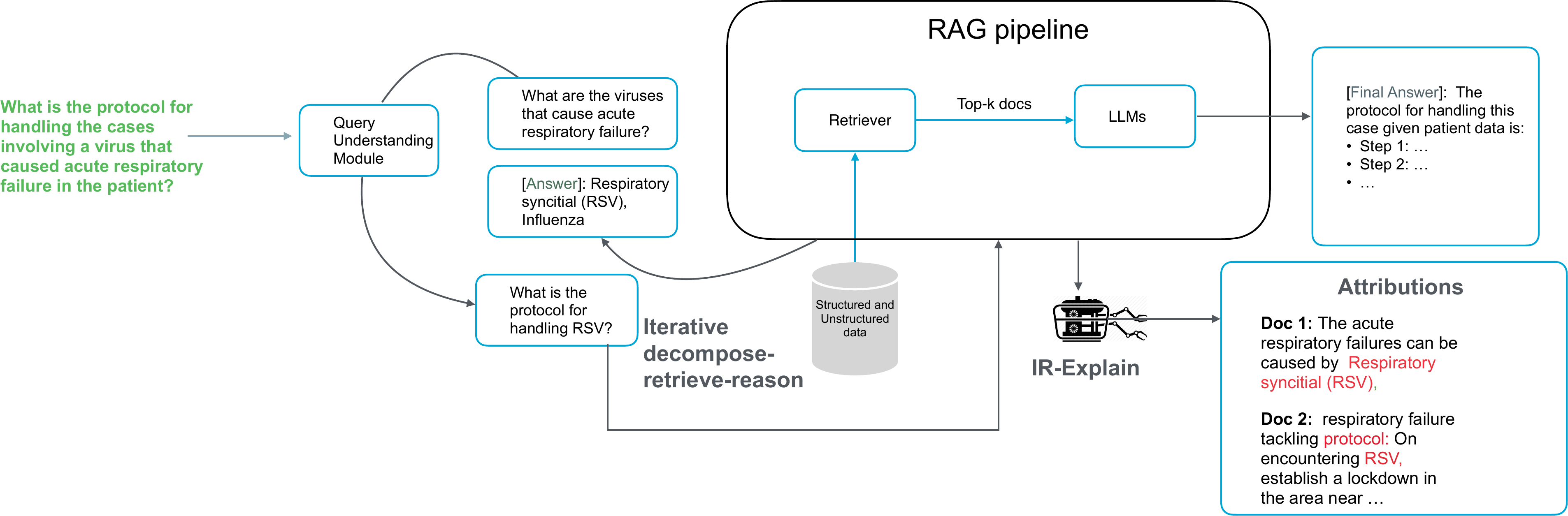}
    \caption{Example of how \irexplain can be used to detect salient terms for RAG attributions.}
    \label{fig:rag_attributions}
\end{figure*}

\begin{table}[tb]
    \centering
    \caption{Content of a document and the perturbed variant ($D'$) of it. New words added to $D'$ are shown in bold, modified words are underlined, and words that are removed are struck out.}
    \label{table-content}
    \fontsize{6pt}{7pt}\selectfont
    \renewcommand{\tabcolsep}{2pt}
    \renewcommand{\arraystretch}{1}
    \begin{tabular}{@{}clll@{}}
        \toprule
        \multicolumn{4}{c@{}}{\textit{Query (qid: 1112341) what is the daily life of thai people }} \\
        \midrule
        Rank & Document & Rel. & Content \\
        \midrule
        1 & $D$ ({docid: 8139255}) & 3 &  An important thing in everyday life is SANUK. Thai people love to \\  
        &  & & have fun together. SANUK can represent many things : eat together, \\ 
        & & & to be with friends and chat, to go out with friends. For Thai people \\
        & & &  SANUK happens with several persons. \\
        \midrule
        - & $D'$ (Perturbed) & - &  An important thing in everyday life is SANUK. Thai people love to \\
        &  & &  have fun together. SANUK \st{can} \underline{represents} many things : eat together, \\
        &  & &  to be with friends and chat, to go out with friends. For \textbf{most} Thai \\
        &  & & people SANUK happens with \underline{multiple} persons. \\
        \bottomrule
    \end{tabular}
\end{table}

\begin{figure}
	\centering
	\begin{subfigure}{0.47\linewidth}
		\includegraphics[width=\linewidth]{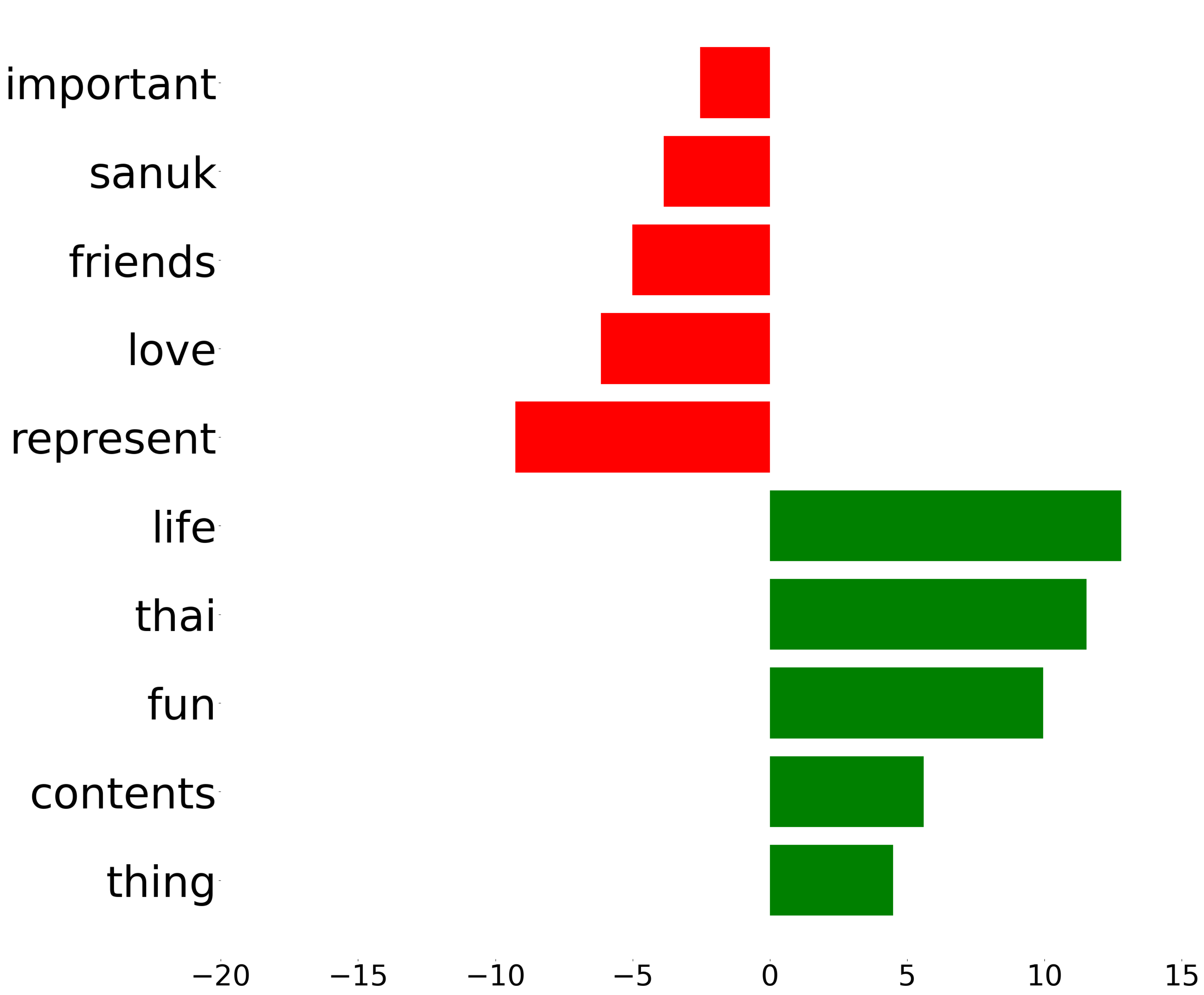}
		\caption{Original document $D$ (\module{docid: 8139255})}
		\label{fig:subfigA}
	\end{subfigure}
	\begin{subfigure}{0.47\linewidth}
		\includegraphics[width=\linewidth]{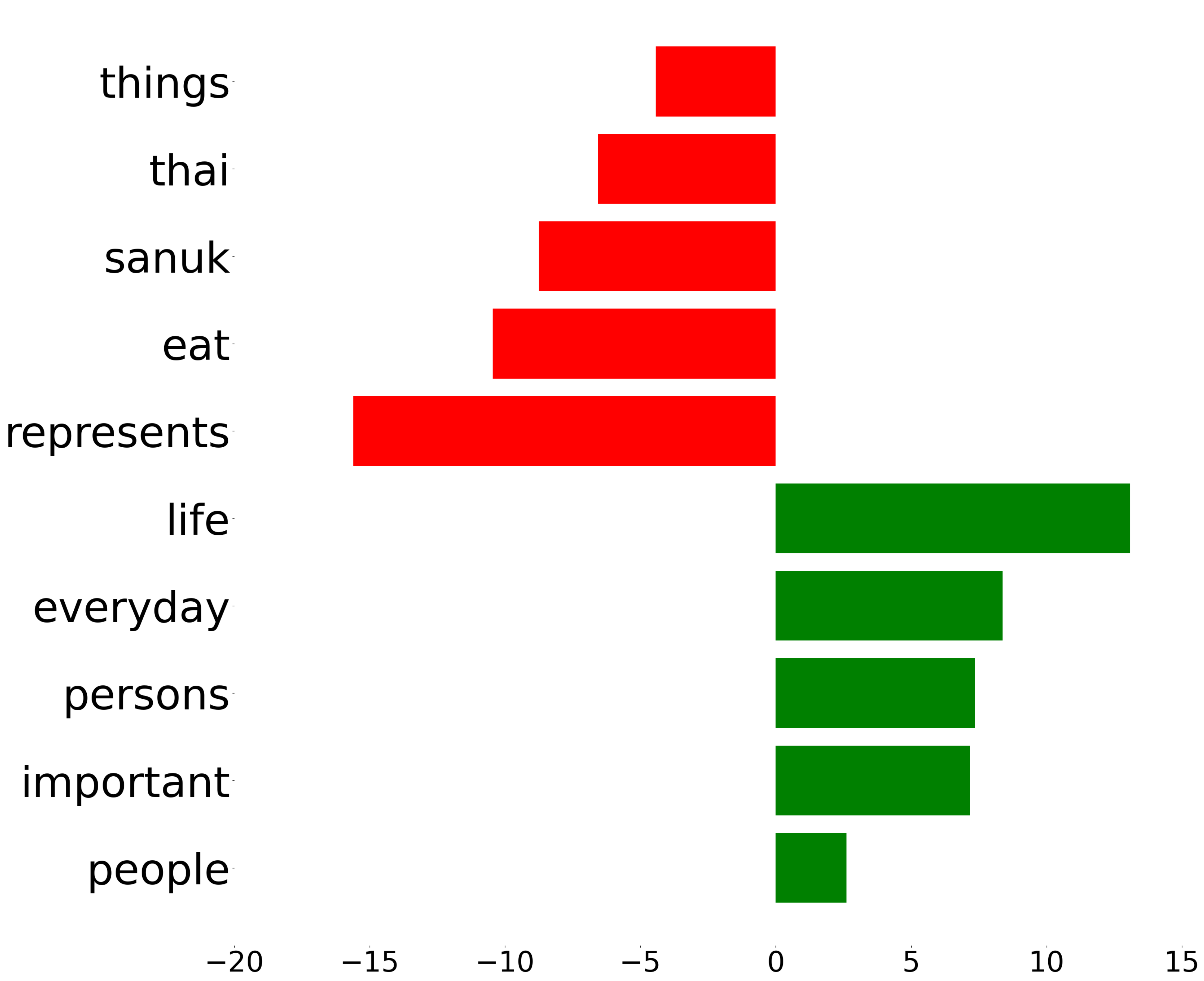}
		\caption{Perturbed document $D'$ (from document $D$)}
		\label{fig:subfigB}
	\end{subfigure}
	\caption{
 Top-$5$ explanations terms of a sample document $D$ and a slightly perturbed instance of $D$.
 The content of both the documents are shown in Table~\ref{table-content}. We show the top-$5$ explanation terms, responsible for positively and negatively influencing the score of these documents. 
    X-axis shows the magnitude of these terms and the Y-axis plots the explanation terms for the query \emph{`what is the daily life of thai people' (TREC DL hard qid: 1112341)}.}
	\label{fig:subfigures}
\end{figure}
\section{Demonstration of Use Cases}
\label{sec:exp}
As an all-in-one toolkit for ExIR methods, \irexplain can be used
for various experiments and analyses.
We demonstrate few use cases of \irexplain
here: (i) robustness analysis of pointwise explanations, (ii)
investigating whether listwise explanations are
sufficiently intuitive, (iii) usecase of pairwise explanation component, (iv) reproducibility study with \irexplain, (v) demonstration with RAG pipeline.  
We have used the MS MARCO passage, document collections, the TREC DL 2019,
2020, and hard~\cite{mackie2021dlhard} topic sets to articulate these use
cases. 
We show that interesting insights and diverse analysis can be obtained easily using \irexplain. We believe this package will inspire and aid further research in ExIR.

\begin{table*}[t]
    \setlength{\tabcolsep}{3pt}
    \renewcommand{\arraystretch}{1.}
    \caption{A subset of results obtained by replicating experiments with
      BFS and Greedy based listwise explainers on TREC 2019 topic set.
      MAP\textsubscript{(rep.)} and RBO\textsubscript{(rep.)} represent our
      obtained values of MAP and RBO respectively (RBO values were computed using
      $p=0.9$). IR models are
      grouped into sparse retrievals (Sparse), re-ranking-based dense
      retrieval (Rerank), and end-to-end dense (E2E) models. These three
      categories of models are similar to that
      in~\cite{llordes-sigir-23}.} 
    \label{table-reproducibility-explain-like-bm25}
    \begin{tabular*}{\textwidth}{@{\extracolsep{\fill}}l@{\qquad\quad}cc@{\qquad\quad}cccc@{\qquad\quad}cccc@{}}

        \toprule
        \textbf{Model} &  &  & \multicolumn{4}{@{}c@{\qquad\quad}}{\module{Greedy}} & \multicolumn{4}{@{}c@{}}{\module{BFS}}\\

        \cmidrule(r{27pt}){4-7}  \cmidrule{8-11} 
            \textbf{Type} & IR Model & MAP & MAP & MAP\textsubscript{(rep.)} & RBO & RBO\textsubscript{(rep.)} & MAP & MAP\textsubscript{(rep.)} & RBO & RBO\textsubscript{(rep.)}  \\
        \midrule
        \multirow{ 2}{*}{Sparse} & BM25 & 0.1067 & - & - & - & - & - & - & - & -  \\
         & RM3 &  0.1411 & - & - & - & - & - & - & - & -  \\
        \midrule
        \multirow{ 2}{*}{Rerank}  & DCT  & 0.2192 & 0.1544 & 0.1444 & 0.2207 & 0.2170 & 0.2050 & 0.1922 & 0.4946 & 0.4055 \\
        & QE\textsubscript{BERT}  & 0.2199  & 0.1414  & 0.1382 & 0.2213 & 0.2208 & 0.2065 & 0.1852 & 0.5015 & 0.4107\\
        \midrule
        \multirow{3}{*}{E2E} & ANCE & 0.1836  & 0.1454 & 0.1393 & 0.2239 & 0.1982 & 0.1723 & 0.1744 & 0.4969 & 0.3815\\
        & CBERT &  0.2182 & 0.1506  & 0.1429 & 0.2230   & 0.2093 & 0.2064 & 0.1846 & 0.4888 & 0.4012\\
        & MonoT5 &  0.2184 & 0.1470   & 0.1502 & 0.2224 & 0.2473 & 0.1920 & 0.1972 & 0.5194 & 0.4215\\
        
        \bottomrule

    \end{tabular*}
\end{table*}

\subsection{Robustness of Pointwise Explanation}
\label{sec:robustness}
This section presents an anecdotal analysis of the robustness of the pointwise explainer \module{EXS}. Specifically, we examine the following hypothesis:
if two documents $D$ and $D'$ are very similar in their content, the pointwise explanations 
provided for them by an explainer 
should not differ much. As an example, we consider the query \emph{`what is the daily life of thai people'} (qid: 1112341) in the MS MARCO passage collection. For this query, 
BM25 retrieves a document $D$ (docid: 8139255, judged `highly relevant') at the 5th position; reranking with SBERT~\cite{sbert}
it to the top position.
We use \module{EXS} from \irexplain to generate a pointwise explanation for $D$. We also construct $D'$ by perturbing $D$ slightly, without altering the actual intent of the document at all. Table~\ref{table-content} shows the contents of both the documents. 
Figure~\ref{fig:subfigures} shows the explanation terms and weights produced by \module{EXS} for both $D$ and $D'$.
The first striking observation is that the explanation terms for $D$ and
$D'$ differ a lot. Out of 10 terms shown for each document, only 3 are
common, namely \module{represent/represents}, \module{sanuk} and
\module{important}. According to \module{EXS}, two versions of the same
term (\module{represents} in $D$, and \module{represent} in $D'$) influence the rankings very negatively. On the other hand, the term \module{important} is said to have a positive influence on $D$ whereas its influence on $D'$ is negative.  
Another observation is that \module{EXS} marks \module{`sanuk'} as a negatively influencing term in both instances. However, if we look at the content of the document (Table~\ref{table-content}), \module{`sanuk'} means ``having fun''; from a user's point of view, it is the most important word for this document. 
Clearly, the explanations provided by \module{EXS} are not stable for this example. 
While one example does not prove that a particular explainer is not robust,
it does show \irexplain can be used easily for a more detailed analysis of
explainers, that leads to interesting observations.




\subsection{Approximating a Ranking as Explanation?}
\label{sec:concepts}
A listwise explainer claims to have explained a complex model $M_2$ if the
ranked lists produced by $M_2$ and the explanation $\langle Q_{\mathit{exp}},
\mathit{SM} \rangle$ are `close enough' according to some measure. However, the
primary objective of generating \emph{human-understandable} explanations
gets overlooked if the success of an explainer is evaluated only in terms
of rank correlation measures. Even when $\langle Q_{\mathit{exp}}, \mathit{SM} \rangle$
closely approximates $M_2$, the question remains: do the terms in
$Q_{\mathit{exp}}$ provide an intuitive explanation to the user? We show
examples of listwise explanations provided by
different approaches for two queries, and look at how well they relate intuitively to the
query.
Table~\ref{table-explanation-terms} shows the explanations generated by
BFS, Greedy, Multiplex, and IntentEXS
explainers for two queries from the TREC DL hard query set when
TCT-ColBERT~\cite{tct_colbert} is used as the black-box neural ranking
model. All of them approximate the ranking of TCT-ColBERT well, but for the
first example (\emph{`what is the daily life of thai people'}) the
explanation terms generated by \module{IntentEXS} include many 
unrelated terms. A similar observation holds for \module{Greedy} in the
second example (\emph{`causes of stroke?'}): most terms do
not convey the underlying intent of the 
query. Thus, from a user's point of view, these explanations are not
intuitive, as they fail to explain the intent of the query.
This highlights the need for a more carefully designed protocol for
evaluating explanation strategies. 
\begin{table}[ht]
    \centering
    \caption{Sample output of four different listwise explanation approaches 
    generated on TREC DL hard queries. MSMARCO passage collection is considered for the retrieval. }
    \label{table-explanation-terms}
    \fontsize{7pt}{8pt}\selectfont
    \renewcommand{\tabcolsep}{5pt}
    \renewcommand{\arraystretch}{1.2}
    \begin{tabular}{@{}cl@{}}
        \toprule
        \textbf{Explainer} & \textbf{Explanation terms} \\
        \midrule
        \multicolumn{2}{l@{}}{\textit{Query: what is the daily life of thai people (qid: 1112341)}} \\
        \midrule
        \module{BFS} &  `thai', `sanuk', `temper', `everydai', `life' \\
        \texttt{Greedy} &  `sanuk', `chat', `friend', `thing', `togeth', \\
                        &  `happen', `everydai', `thai', `life', `fun' \\
        \texttt{Multiplex} &  `buddhist', `thailand', `85,000', `anger', `life' \\  
                        & `daily', `monk', `people', `what'\\
        \texttt{IntentEXS} &  `thai', `6.00', `peopl', `life', `µà', `put', `though',  \\
                           &  `do', `vast', `friend' \\
        \midrule
        \multicolumn{2}{l@{}}{\textit{Query: causes of stroke? (qid: 88495)}} \\
        \midrule
        \module{BFS} &  `causes', `stroke', `high', `arteri' \\
        \texttt{Greedy} &  `uncommon', `vast', `harden', `rarer', `less', \\
                        &  `list', `well', `mani', `follow', `stroke' \\
        \texttt{Multiplex} &  `stroke', `clot',  `causes', `vast', `die'\\
        \texttt{IntentEXS} &  `brain', `arteri', `list', `usual' \\
        \bottomrule
    \end{tabular}
\end{table}

\subsection{Usecase of Pairwise Explanation}
Recall from Section~\ref{sec:list-wise} that Multiplex and IntentEXS are designed to optimize for maximum preference pairs ($\mathcal{L}$). The output of these explainers, i.e., the explanation terms they produce, can also be used to provide a pairwise explanation induced by $D_j$ and $D_k$. These explanations can be easily visualized in \irexplain by invoking the \module{show\_matrix()} function of these listwise components and examining at the specific column of $\langle D_j, D_k \rangle$.

\subsection{Replicability Study}
With our library, replicating\footnote{We use this term and its variants in
accordance with the definitions provided in
\url{https://www.acm.org/publications/policies/artifact-review-and-badging-current}.}
baselines becomes a much easier task. As an example, we replicated the
results produced by~\citet{llordes-sigir-23} using \irexplain.
Table~\ref{table-reproducibility-explain-like-bm25} shows the results
obtained using BFS and Greedy approaches on the TREC 2019 query set. The
experiments were conducted on the same set of IR models (DCT~\cite{dct},
QE\textsubscript{BERT}~\cite{llordes-sigir-23}, ANCE~\cite{ance},
CBERT~\cite{khattab:2020:sigir:colbert}, MonoT5~\cite{monot5}) as
in~\cite{llordes-sigir-23}. We present a subset of the results and omit the
Jaccard-based overlap measure; instead, we report RBO ($p=0.9$), which appears to be
preferred as a fidelity measure in recent literature.
Obtained versions of MAP and RBO are denoted as MAP\textsubscript{(rep.)} and RBO\textsubscript{(rep.)} respectively.
As observed from the table, the trends in the obtained figures are mostly
similar to those reported by \citet{llordes-sigir-23}. However, for
MonoT5-based ranking models, the Greedy baseline performs better than the
reported figures, whereas in other cases, the RBO\textsubscript{(rep.)}
values are smaller. Though the randomness involved in their method is a
potential source of these variations, a more
thorough investigation is needed to determine the cause of this drop in
performance.  
Note that we could not perform significance testing or compute
replicability metrics due to the unavailability of the required details in \citet{llordes-sigir-23}. 

\subsection{Demonstration with RAG Pipeline}

\irexplain would aid in interpreting Retrieval-Augmented Generation (RAG) pipelines. While LLMs have made strides in a wide range of NLP tasks they still suffer from hallucination \cite{hallucination}, where the LLM makes up plausible yet non-factual statements due to knowledge gaps. While RAG \cite{rag} aids in mitigating this issue to an extent the problem of attribution persists. Attribution entails determining if the answer generated by the LLM is faithfully supported by the retrieved documents. This task is difficult due to the large number of possible documents and large output space \cite{qi-etal-2024-model,xu-etal-2023-critical}. While several answer attribution approaches have been proposed \cite{Rashkin2021MeasuringAI,bohnet2023attributedquestionansweringevaluation}, they still rely on an external validator, such as NLI (Natural language Inference) models. hence, such approaches cannot guarantee the faithfulness of the attribution process. Hence, more recently approaches like MIRAGE \cite{qi-etal-2024-model} which leverage the retrieval and LLM model internals for ensuring faithfulness of attributions. MIRAGE detects
context-sensitive tokens in the LLM output and pairs them
with retrieved documents, contributing to their
prediction via saliency methods. \irexplain would be of immense utility in such attribution frameworks to detect salient terms through principled explainability approaches for debugging the retrieval model and the generative outputs. An overview of how \irexplain would aid in attribution for RAG systems is shown in Figure \ref{fig:rag_attributions}.

\section{Summary and Future Work}
\label{sec:concl}
This paper presents a Python library \irexplain implementing all the popular post-hoc approaches in ExIR. The library is integrated with Pyserini and \irdatasets, and integration with PyTerrier is in progress. 
{Users of \irexplain can use several sparse and dense retrieval models and standard test collections for a rigorous analysis of ExIR methods. 
As we have demonstrated via a few use cases, this library would make it convenient for researchers to use, analyse and compare different explanation approaches, and identify the shortcomings of the current approaches. In turn, this should inspire and help future research in ExIR. 
As part of future work, we plan to integrate \irexplain with various visualization tools, and the ongoing probing components.
Rigorous evaluation is another challenge for ExIR, with many studies providing mostly anecdotal evidence. We therefore plan to incorporate a rigorous evaluation framework for ExIR approaches within \irexplain.}